# Intrinsic Quantum Clusters in Kagome Weyl Semimetal $Co_3Sn_2S_2$

Yuqing Xing[1,2]†, Hui Chen[1,2]†, Li Huang[1,2]†, Roger Guzman[2]†, Qi Zheng[1,2], Senhao Lv[1,2], Jinan Shi[2], Haitao Yang[1,2], Wu Zhou[2*], and Hong-Jun Gao[1,2*]

[1]Beijing National Laboratory for Condensed Matter Physics and Institute of Physics, Chinese Academy of Sciences, Beijing 100190, China

[2]School of Physical Sciences, University of Chinese Academy of Sciences, Beijing 100190, China

†These authors contributed equally to this work

*Correspondence to: hjgao@iphy.ac.cn, wuzhou@ucas.ac.cn

# Abstract

Impurities and intrinsic point defects, which profoundly influence spin, charge, and topological degrees of freedom, are crucial parameters for tuning quantum states in quantum materials. The magnetic Weyl semimetal $Co_3Sn_2S_2$ with its strong spin–orbit coupling, intrinsic ferromagnetism, and kagome lattice of correlated electrons, provides a compelling platform for studying impurity excited states. Yet, the role of intrinsic impurities in shaping its quantum states remains elusive. Here, we uncover intrinsic quantum clusters—localized intrinsic point defects that act as tunable quantum perturbations capable of reshaping electronic states and order parameters, on the surface of $Co_3Sn_2S_2$ via scanning tunneling microscopy/spectroscopy and non-contact atomic force microscopy, combined with scanning transmission electron microscopy–electron energy loss spectroscopy. These clusters are identified as native oxygen defects that dominate the intrinsic defect landscape on both cleaved surface terminations. On the Sn-terminated surface, oxygen impurities occupy hollow sites between three Sn atoms, and tune the flat band near the Fermi level, which exhibits orbital magnetism induced unconventional Zeeman effect under an applied magnetic field. On the S-terminated surface, oxygen interstitials reside slightly off-center relative to the S lattice and generate occupied impurity states that retain sixfold symmetry at higher energies but reduce to $C_2$ symmetry at lower energies. In contrast, these impurity states show no measurable magnetic response. Our findings establish that intrinsic oxygen-related quantum clusters act as tunable local perturbations in a topological kagome magnet, offering a versatile platform to probe and engineer impurity-driven phenomena in correlated and topological systems.

**Main Text**

Impurity is a key parameter for emerging and tuning exotic quantum states in the host materials[1–3]. In semiconductors, notably, a charge neutral vacancy can capture an electron or emit an electron to become charge states, therefore acting as a dopant that modifies carrier concentration and local electrostatics[4]. Understanding defect behaviors constitutes a significant step in defect engineering that not only modifies the conductance of semiconductor but also enables novel material functions such as high-spin states of nitrogen-vacancy centers in diamond[4] and ionic diffusion paths enables by oxygen vacancies in perovskites[5]. Beyond conventional semiconductors, vacancies and impurity defects also play crucial roles in quantum materials, where real-space disorder can strongly influence spin, charge, and topological degrees of freedom[1]. In particular, localized intrinsic point defects—termed intrinsic quantum clusters— act as tunable quantum perturbations capable of reshaping electronic states and order parameters. Among these materials, topological semimetals like $Co_3Sn_2S_2$ have emerged as a representative magnetic Weyl semimetal that combines intrinsic ferromagnetism, strong spin-orbit coupling, and a kagome lattice of correlated Co $d$ electrons[6–11]. These ingredients give rise to an unusual topological band structure featuring flat bands, surface Fermi arcs and manifest in a range of exotic phenomena such as the giant anomalous Hall[6] and Nernst effects[12], orbital magnetism[13], chiral phonons[14] and magnons[15]. Understanding how intrinsic quantum clusters govern these emergent behaviors in $Co_3Sn_2S_2$ is thus crucial both for revealing their microscopic mechanisms and for accessing the intrinsic quantum limit.

Experimental studies on impurities in $Co_3Sn_2S_2$ have made significant progress in recent years. The introduction of Fe and Ni impurities can effectively tune the magnetic properties of $Co_3Sn_2S_2$[16,17]. It was reported that by introducing non-magnetic In doping in $Co_3Sn_2S_2$, the defect would establish a spin-polarized quantum impurity state with anomalously large orbital magnetic moment[18–20]. In addition, the localized spin-orbit polarons have been observed around the single S vacancies on the S-terminated surface[21]. The atomic manipulation of S vacancies and the spin-orbit polaron has also been achieved recently, emphasizing the tunable electronic and magnetic properties and the ability of engineering specific vacancy-induced bound states at atomic scale[22]. The chemical bonding in adatom arrays has been studied by using machine learning[23]. Very recently, ultra-high-quality $Co_3Sn_2S_2$ crystal has been synthesized and exhibits gigantic anomalous Hall conductivity and exceptional carrier mobility, which is much higher than previous work of pristine crystal[24]. These advances strongly motivate exploration of how intrinsic quantum clusters influence spin–orbit coupling and the local electronic structure in this topological kagome magnet.

Here we investigate the structural configurations and electronic states of intrinsic quantum clusters in $Co_3Sn_2S_2$, by combining scanning tunneling microscopy/spectroscopy (STM/STS), non-contact atomic force microscopy (nc-AFM), and scanning transmission electron microscopy–electron energy loss spectroscopy (STEM–EELS). We identify these clusters in the magnetic Weyl semimetal $Co_3Sn_2S_2$ as native oxygen defects. On the Sn-terminated plane, intrinsic impurities locate in the threefold hollow sites among Sn atoms and perturb the flat band near the Fermi level, giving rise to a pronounced, linear energy shift under ±8 T field-a clear signature of orbital magnetism. In contrast, on the S termination the impurity sits slightly off the lattice site and produces in-gap states that retain $C_6$ symmetry at higher energies but collapse to $C_2$ at lower energies, while remaining completely insensitive to magnetic field.

$Co_3Sn_2S_2$ has a layered structure in which kagome $Co_3Sn$ layer is sandwiched between hexagonal S and Sn layer (Fig. 1a-b). Single crystals of $Co_3Sn_2S_2$ were cleaved *in-situ* in ultra-high vacuum at low temperature of about 10 K. After cleavage, two terminated surfaces are commonly observed, which are previous identified to be Sn-terminated surface and S-terminated surface[7,11,25]. Therefore, we focus primarily on intrinsic quantum clusters on the Sn and S surfaces.

The randomly-distributed intrinsic impurities exhibit protrusions in the STM topographic image of Sn surface (Fig. 1c). Zoom-in topography (Fig. 1d) of the protrusion exhibits a slightly off-centered triangle shape intensity distribution in between three surface Sn atoms. As it is difficult to identify the type and exact position of defect in STM measurements which are sensitive to the electronic states, we applied nc-AFM, by probing the atomic force gradient in frequency modulation mode, to directly image the atomic positions (the core electrons)[26,27]. The nc-AFM can provide important supplement information on topographically or chemically induced features, such as defects and impurities, which usually cannot be resolved clearly by STM[28,29]. For the bright triangle in the STM image of Sn surface (Fig. 1e), the AFM image reveal a bright protrusion (dashed blue circle in Fig. 1e), in the center hollow site of hexagonal lattice of Sn-terminated surface and all Sn atoms are occupied without vacancies. The bright protrusion is 15 pm higher than typical Sn atoms. In addition, one neighboring atom in the Sn lattice site is 20 pm higher than other five neighboring atoms. Therefore, the round protrusion in STM image of S surface are attributed to a state of single impurity which occupy at the hollow site of the S hexagonal lattice and may possibly be bound to one of the neighboring three Sn atoms.

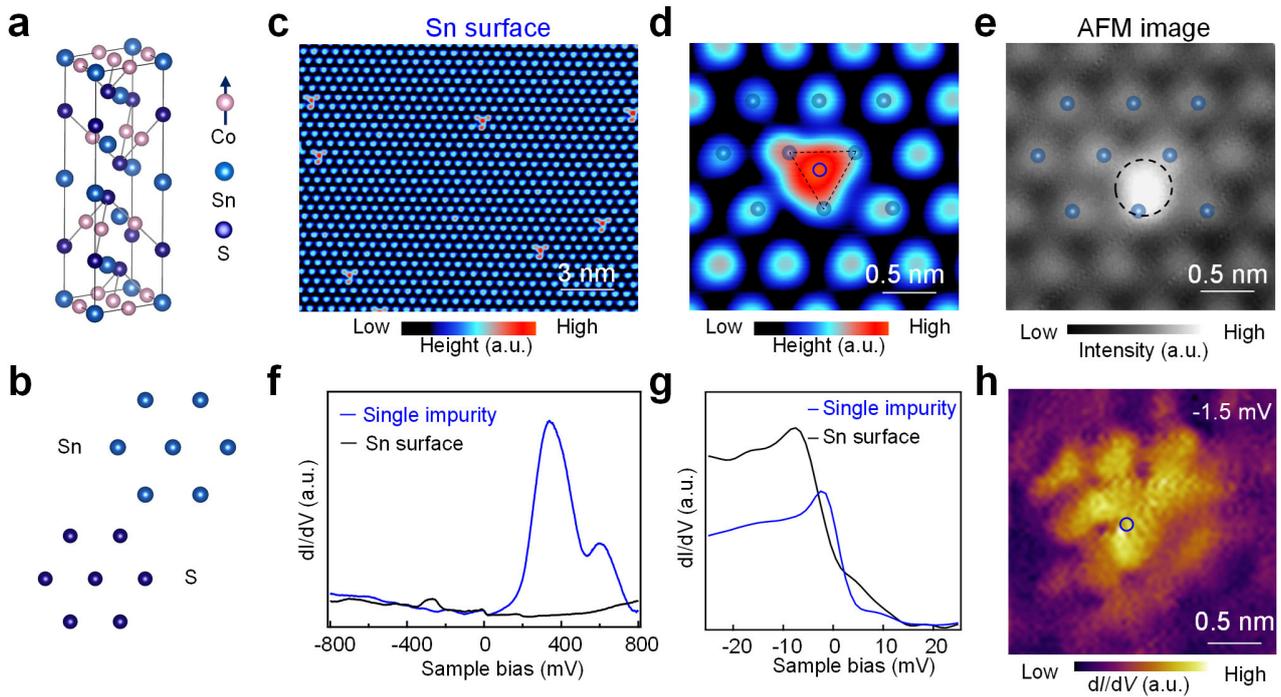

**Fig. 1. STM topography, d$I$/d$V$ spectra and nc-AFM images of intrinsic impurities on Sn terminated surface.** (a) Schematics showing atomic structure of Co$_3$Sn$_2$S$_2$. (b) Schematics showing atomic structure of Sn terminated layer (top) and hexagonal S terminated layer. (c) Large-scale STM topographic image of Sn terminated surface of Co$_3$Sn$_2$S$_2$, showing randomly distributed round-shaped impurities. ($V_s$ = −400 mV, $I_t$ = 1 nA) (d) Zoom-in STM image of a round impurity, illustrating its asymmetric hybridization with neighboring atoms. The blue circle highlights the central position of the impurity. ($V_s$ = −100 mV, $I_t$ = 500 pA) (e) Non-contact AFM (nc-AFM) images of the same single impurity on Sn termination, overlapped by the illustrated atomic image of Sn layer. The black circle highlights the central position of the impurity from the AFM image. ($V_{sample}$ = -400 mV, $I$ = 10 pA; $\Delta f$ = -14 Hz). (f) d$I$/d$V$ spectra on (blue curve) and off (black curve) single impurity on the Sn surface, ranging from -800 mV to +800 mV. ($V_s$ = −800 mV, $I_t$ = 500 pA, $V_{mod}$ = 0.2 mV). (g) d$I$/d$V$ spectra on (blue curve) and off (black curve) single impurity on the Sn surface, ranging from -25 mV to +25 mV. ($V_s$ = −200 mV, $I_t$ = 1 nA, $V_{mod}$ = 0.2 mV). (h) Differential conductance map at -1.5 mV, showing a triangular pattern associated with the impurity extended to roughly 15 atoms around the impurity ($V_s$ = −-20 mV, $I_t$ = 1 nA, $V_{mod}$ = 0.1 mV).

The intrinsic impurities also induce local electronic states at Sn surface. In larger energy range, a typical single impurity d*I*/d*V* spectrum shows large enhancement above the Fermi energy at 350 meV and 600 meV. As the pristine Sn surface exhibits a flat band induced peak near the Fermi energy[7], it is also important to check if the impurity perturbs this spectra feature. As in Fig. 1g, such peak moves towards positive bias for ~5 meV on the single impurity compared to that of impurity-free Sn surface, suggesting a local hole-dope behavior. Additionally, a d*I*/d*V* map at -1.5 mV shows a distribution that extends to 15 surrounding atoms nearby, also suggesting its chemical doping behavior to the itinerant electron near the Fermi energy.

On the other termination, composed of sulfur atoms arranged in a hexagonal lattice, the surface exhibits a vacancy-rich morphology[7,11,21,22,25]. Similar as on Sn-terminated surface, randomly distributed impurities are observed (Fig. 2a). Notably, the overall impurity density is comparable to that reported in earlier studies involving In-doped impurities on the same S-terminated surface[19]. A close-up view of single impurity (Fig. 2b) reveals an atomically sharp and hexagonally-symmetric protrusion. To further explore the atomic origin of these defects, non-contact AFM imaging was performed (Fig. 2e), revealing that the bright electronic state in STM image (Fig. 2b) arises from an atom off center from the S lattice, as marked by the white circle in fig. 2c, with the underlying S atom highlighted in red. The atom is 80 pm off center along the crystal direction and 1.5 pm lower than other S atoms in the AFM image.

Spectroscopic measurements (Fig. 2d) taken directly on a single impurity show pronounced in-gap features, with localized states emerging from the occupied side of the spectrum below –200 mV, in contrast to the relatively featureless gap-like spectrum taken off the defect site. Notably, a spatial map of the differential conductance at the energy of localized state (e.g. –264 mV in Fig. 2e) shows a sixfold (C6) symmetric pattern centered at the impurity (marked by orange dotted hexagon in Fig. 2e), indicating a strong coupling between the impurity state and the underlying lattice symmetry. Interestingly, a d*I*/d*V* map at –362 mV (Fig. 2f) reveals symmetry lowering of the defect state from C6 to C2, similar to the AFM image in Fig. 2c, suggesting a potential energy-dependent orbital or structural anisotropy associated with the impurity.

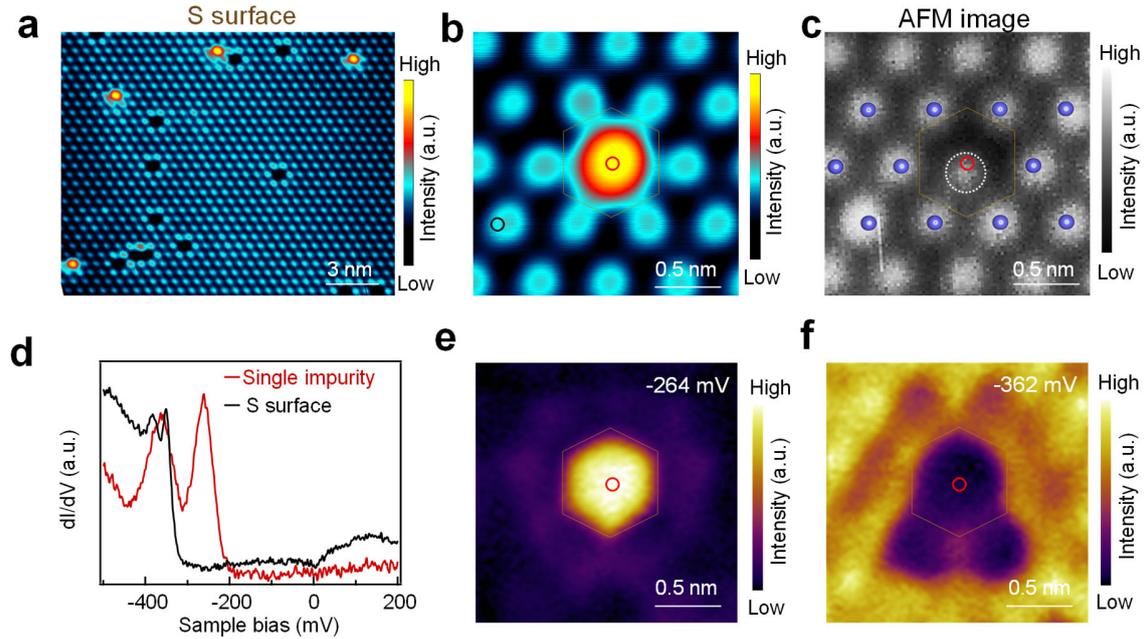

**Fig. 2. STM topography, d$I$/d$V$ spectra and nc-AFM images of intrinsic impurities on S terminated surface.** (a) Large-scale STM topographic image of S terminated surface, identified by characteristic S vacancies, showing randomly distributed round-shaped impurities roughly on top of the S atomic site. ($V_s$ = −400 mV, $I_t$ = 500 pA). (b) Zoom-in STM image of the impurity. ($V_s$ = −400 mV, $I_t$ = 500 pA). (c) nc-AFM images of the same type of single impurity on S termination, overlapped by the illustrated atomic image of S layer. The white dashed circle highlights the position of the impurity, while the red circle indicating the location of the covered S atom. ($V_{sample}$ = -400 mV, $I$ = 10 pA; $\triangle f$ = -9 Hz). (d) d$I$/d$V$ spectra on (red curve) and off (black curve) single impurity on the S surface, ranging from -500 mV to +200 mV, showing distinctive defect states emerging from the negative energy side. ($V_s$ = −500 mV, $I_t$ = 500 pA, $V_{mod}$ = 0.2 mV). (d) Differential conductance map at -264 mV of the impurity, showing C6 symmetric pattern centered around the surface impurity. (f) dI/dV map slice at -362 mV of the same impurity. The symmetry of the pattern reduces from C6 to C2, indicating a different origin of the defect states compared to those at -264 mV. STM parameters for (e-f): $V_s$ = −400 mV, $I_t$ = 500 pA, $V_{mod}$ = 0.2 mV.

The intrinsic impurities at both Sn and S surface exhibit off-center structures, indicating that it may result from the intrinsic interstitial impurity. To probe the chemical nature of interstitial impurities, we conducted atomic-resolution scanning transmission electron microscopy (STEM) combined with electron energy loss spectroscopy (EELS). Fig. 3a shows high-angle annular dark field (HAADF) and annular bright field (ABF) images acquired along the [100] zone axis, revealing a localized impurity site residing adjacent to the Sn atomic plane. The corresponding structural model (Fig. 3b) suggests the scenario of an O atom bonding in the vicinity of the Sn plane. EELS measurements at the same region (Fig. 3c) reveal a clear oxygen K-edge onset at ~532 eV in the summed spectra of the impurity site (green), compared to control spectra lacking the oxygen signal (orange). The differential spectrum (black) confirms the presence of a localized oxygen signal and its spatial correlation with the defect.

A second type of impurity configuration is observed in a different layer, where the oxygen atom is situated near the Sn plane (Fig. 3d). HAADF and ABF images again identify the impurity through enhanced contrast in the ABF channel. The corresponding atomic model (Fig. 3e) places the oxygen impurity between S and Co3Sn atoms, introducing different local chemical environments compared to the Sn-adjacent configuration. EELS analysis of this configuration (Fig. 3f) again reveals distinct oxygen-related fine structure features, with a similar onset energy but different spectral shape, indicating a modified bonding environment. The differential spectrum highlights the contrast between O-rich and O-free regions, demonstrating the presence of chemically distinct oxygen impurities occupying inequivalent sites within the lattice. These findings confirm oxygen as a dominant impurity species and provide a microscopic basis for the diverse electronic signatures observed in STM.

We then systematically investigate the evolution of electronic states of two oxygen impurities with their spatial separations on both S- and Sn-terminated surfaces. As shown in Fig. 4a, STM topographies capture two nearby oxygen impurities on the S termination, with varying separation distances that allow us to tune their mutual interaction. Differential conductance spectra (Fig. 4b) taken at the impurity centers (marked by circles) reveal that as the spacing decreases, the defect-induced states progressively broaden and shift in energy, indicating hybridization or interference between their localized states. To assess the potential magnetic nature of these states, we measured the d$I$/d$V$ spectra under perpendicular magnetic fields ranging from –8 T to +8 T (Fig. 4c). The defect states on the S termination exhibit negligible Zeeman splitting or energy shifts across the entire field range, suggesting a non-magnetic origin for these states, making it behave differently comparing to that of the In impurity on the same termination[19].

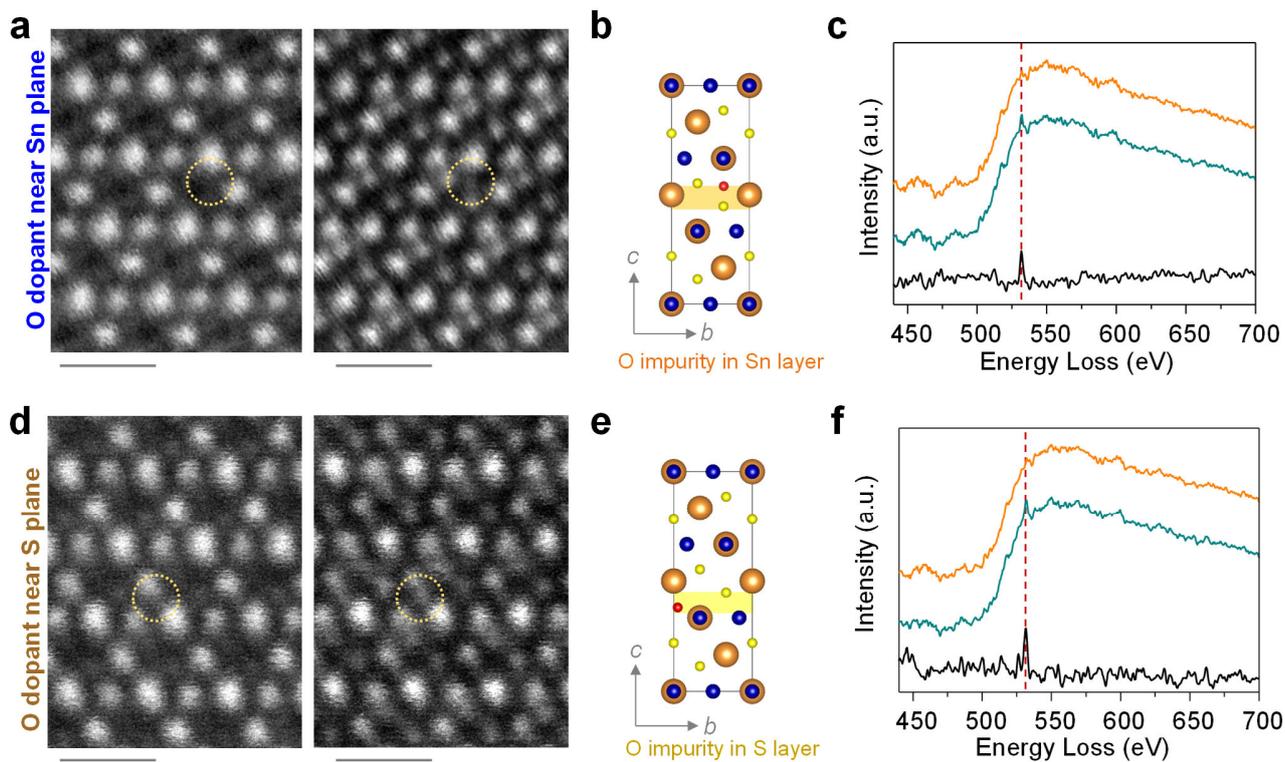

**Fig. 3. Cross sectional STEM-EELS analysis of two types of impurities.** (a) STEM-HAADF (left) and inverted-contrast ABF (right) images of an oxygen impurity located adjacent to the Sn plane (dashed circles) viewed along the [100] zone axis. (b) Structure model of the $Co_3Sn_2S_2$ with an oxygen impurity (red atom) adjacent to the Sn plane. The Co, Sn and S atoms are shown in blue, golden and yellow, respectively. (c) EELS spectra of the Sn M-edge energy loss range showing the summed spectra of the Sn sites with (green line) and without (orange line) oxygen signals at 532 eV, which corresponds to the onset of the O K-edge. The black line is the difference signal between the normalized green and orange spectra. (d) STEM-HAADF (left) and inverted-contrast ABF (right) images of an oxygen impurity located adjacent to the S plane (dashed circles) viewed along the [100] zone axis. (e) Structure model of the $Co_3Sn_2S_2$ with an oxygen impurity (red atom) adjacent to the S plane. (f) EELS spectra of the Sn M-edge energy loss range showing the summed spectra of sites located between pure Sn, $Co_3Sn$ and pure Co columns with (green line) and without (orange line) O fine structure feature. The black line is the difference signal. Scale bars: 0.5 nm.

In contrast, similar experiments on the Sn-terminated surface reveal a different behavior. As shown in Fig. 4d, O impurities are again found with variable spacing, and dI/dV spectra at the impurity sites (Fig. 4e) show sharp peaks in the local density of states close to the Fermi level. These peaks exhibit notable spectral evolution with impurity proximity, consistent with electronic coupling or potential charge redistribution in the vicinity of the Sn layer. As the distance between the two impurities decreases, the doping-induced resonance is quenched, and the peak shifts back to the intrinsic flat-band energy which is ~ -6 mV. Strikingly, when a magnetic field perpendicular to the surface is applied(Fig. 4f), the peak near the Fermi energy exhibits a clear linear shift towards higher energy with increasing field magnitude. This behavior suggests a magnetic-field-sensitive defect state, possibly reflecting orbital magnetism contributions specific to the Sn layer. These distinct magnetic responses highlight how the local atomic environment, specifically the lattice symmetry and bonding configuration strongly influences the nature of the impurity states.

While native defects are often considered undesirable in functional materials, here we show that even unintentional oxygen impurities exhibit rich and tunable behavior—manifesting as spatially localized defect states with distinct spectral, structural, and magnetic-field responses depending on their lattice registry. The comparative analysis of O impurities on Sn- and S-terminated surfaces reveals a striking contrast. On the Sn termination, impurities occupy hollow sites between three Sn atoms and tune the sharp electronic resonances near the Fermi level. These states exhibit sensitivity to inter-impurity spacing and show a pronounced linear shift under applied magnetic field, indicative of an orbital magnetism dominated state. This behavior echoes the previously reported atomically engineered quantum states[22], although the oxygen-derived states here arise intrinsically and occupy distinct lattice environments. On the S termination, in contrast, oxygen impurities reside slightly off-center from the S atomic lattice, as directly visualized by nc-AFM. They introduce in-gap states that preserve the sixfold symmetry of the underlying lattice at certain energies but break it at lower energy, indicating an energy-dependent orbital anisotropy. Importantly, these states exhibit no discernible response to magnetic field, suggesting a nonmagnetic origin. These findings establish that even nominally intrinsic oxygen impurities in $Co_3Sn_2S_2$ behave as intrinsic quantum clusters, acting as local quantum perturbations that influence orbital symmetry, hybridization, and potentially magnetism. Their sensitivity to lattice environment, spacing, and magnetic field offers a versatile route for probing and engineering quantum impurity phenomena in topological kagome magnets.

In summary, we investigated oxygen-induced quantum clusters in $Co_3Sn_2S_2$ using multiple approaches combining STM, nc-AFM, and STEM-EELS. Our results identify oxygen as a dominant intrinsic impurity species and reveal two structurally and spectroscopically distinct configurations depending on whether the impurity resides on the S- or Sn- termination. These impurities generate spatially localized electronic states with distinct symmetry and different magnetic field response behavior. The interplay between impurity location, orbital anisotropy, and topological band structure in this kagome magnet prove the nature of quantum clusters. Our findings provide a new platform to study and manipulate quantum impurity states in a topological context.

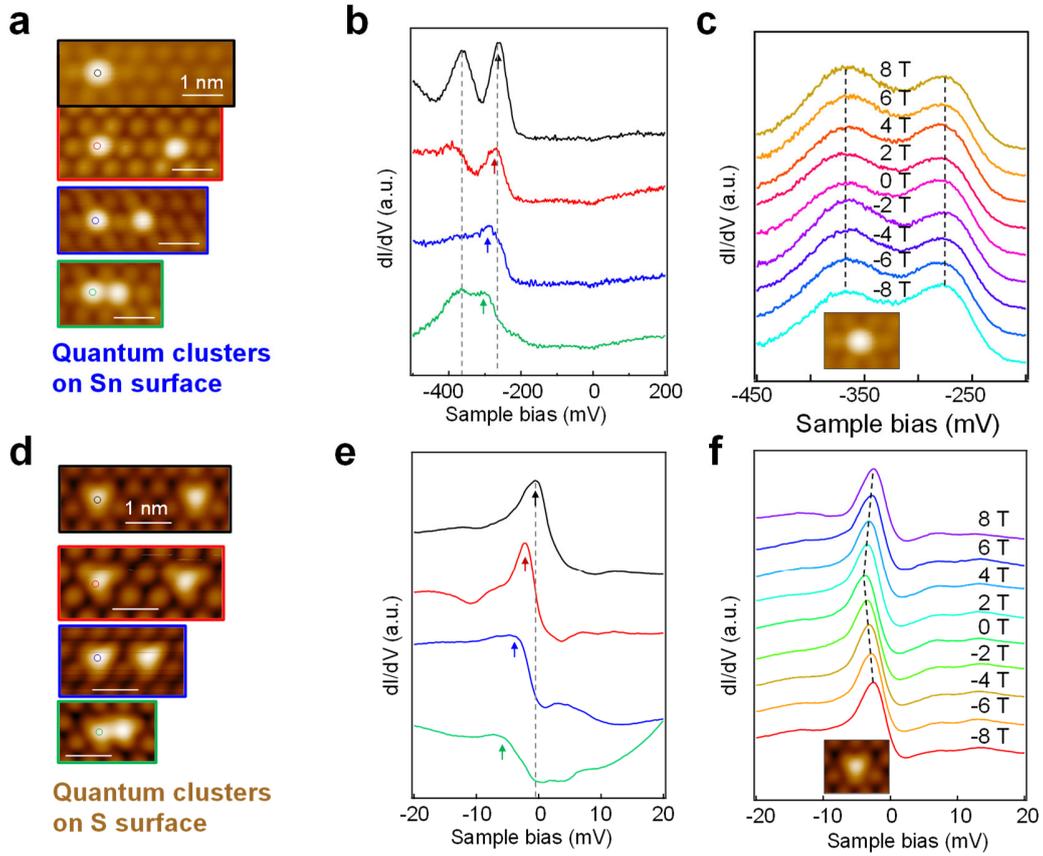

**Fig. 4. Spatial interactions and magnetic field dependent d$I$/d$V$ of oxygen-induced quantum clusters on Sn- and S- terminated surfaces.** (a) Topographic images showing varying spatial distance between two individual O impurities on S termination ($V_s$ = −400 mV, $I_t$ = 100 pA). (b) d$I$/d$V$ spectra acquired at O impurities marked by circles in fig. 4a, showing the evolution of perturbation strength tuned defect states ($V_s$ = −500 mV, $I_t$ = 500 pA, $V_{mod}$ = 0.2 mV). (c) d$I$/d$V$ spectra of the defect states with magnetic field from -8 T to +8 T, showing non-magnetic behavior under the perpendicular magnetic field. ($V_s$ = −450 mV, $I_t$ = 500 pA, $V_{mod}$ = 0.05 mV). (d) Topographic images showing varying spatial distance between two individual O impurities on Sn termination ($V_s$ = −700 mV, $I_t$ = 100 pA). (e) d$I$/d$V$ spectra at O impurities marked by circles in fig. 4d, showing the evolution of density of states near Fermi level ($V_s$ = −200 mV, $I_t$ = 2 nA, $V_{mod}$ = 0.05 mV). (f) d$I$/d$V$ spectra of the peak near Fermi energy with magnetic field from -8 T to +8 T, showing a linear shift towards higher energy of the peak under the presence of magnetic field ($V_s$ = −200 mV, $I_t$ = 2 nA, $V_{mod}$ = 0.05 mV).

**Method**

Single crystal growth of $Co_3Sn_2S_2$

The single crystals of $Co_3Sn_2S_2$ were grown by flux method with Sn/Pb mixed flux. The starting materials of Co (99.95% Alfa), Sn (99.999% Alfa), S (99.999% Alfa) and Pb (99.999% Alfa) were mixed in molar ratio of Co : S : Sn : Pb = 12 : 8 : 35 : 45. The mixture was placed in $Al_2O_3$ crucible sealed in a quartz tube. The quartz tube was slowly heated to 673 K over 6 h and kept over 6 h to avoid the heavy loss of sulfur. The quartz tube was further heated to 1323 K over 6 h and kept for 6 h. Then the melt was cooled down slowly to 973 K over 70 h. At 973 K, the flux was removed by rapid decanting and subsequent spinning in a centrifuge. The hexagonal-plate single crystals with diameters of 2 ~ 5 mm are obtained. The composition and phase structure of the crystals were checked by energy-dispersive x-ray spectroscopy and x-ray diffraction, respectively.

Scanning tunneling microscopy/spectroscopy

The samples used in the experiments were cleaved *in situ* at 6 K and immediately transferred to an STM head. Experiments were performed in an ultrahigh vacuum ($1\times10^{-10}$ mbar) ultra-low temperature STM system (40 mK) equipped with 9-2-2 T magnetic field. All the scanning parameter (setpoint voltage and current) of the STM topographic images are listed in the captions of the figures. Unless otherwise noted, the differential conductance ($dI/dV$) spectra were acquired by a standard lock-in amplifier at a modulation frequency of 973.1 Hz. Non-magnetic tungsten tip was fabricated via electrochemical etching and calibrated on a clean Au(111) surface prepared by repeated cycles of sputtering with argon ions and annealing at 500 °C.

Non-contact atomic force microscopy.

All nc-AFM measurements were performed at LHe temperature with the base pressure lower than $2\times10^{-10}$ mbar. The nc-AFM measurements were performed using a commercial qPlus tuning fork sensor[30] with a Pt/Ir tip in frequency modulation mode[31]. The resonance frequency $f_0$ = 29.2 kHz, and stiffness about 1800 N/m. The topological nc-AFM images were acquired under constant frequency shift with an oscillation amplitude $A$ = 100 pm.

Scanning transmission electron microscopy/electron energy loss spectroscopy

STEM and EELS measurements were carried out using a Nion HERMES-100, operated at 100 kV and a probe forming semi-angle of 30 mrad.